\documentclass{elsart}
\usepackage{amssymb,graphicx}
\begin{document}
\begin{frontmatter}
\title{{\em Ab Initio} Yield Curve Dynamics}
\author[mcm,osc]{Raymond J.~Hawkins},
\author[osc]{B.~Roy Frieden} and
\author[con]{Joseph L.~D'Anna}
\address[mcm]{Countrywide Bank \\ 225 West Hillcrest Drive, Thousand Oaks, CA 91360, USA}
\address[osc]{College of Optical Sciences \\ University of Arizona,
Tucson, AZ 85721, USA}
\address[con]{SG Constellation LLC \\ 1221 Avenue of the Americas, New York, NY 10020, USA}
\begin{abstract}
We derive an equation of motion for interest-rate yield curves by
applying a minimum Fisher information variational approach to the
implied probability density.  By construction, solutions to the
equation of motion recover observed bond prices.  More
significantly, the form of the resulting equation explains the
success of the Nelson Siegel approach to fitting static yield curves
and the empirically observed modal structure of yield curves.  A
practical numerical implementation of this equation of motion is
found by using the Karhunen-L\`{o}eve expansion and Galerkin's
method to formulate a reduced-order model of yield curve dynamics.
\end{abstract}
\thanks{We thank Prof.~Ewan Wright for helpful discussions and encouragement.}
\begin{keyword}
Bond, interest rate, dynamics, Fisher information, yield curve, term
structure, principal-component analysis, proper orthogonal
decomposition, Karhunen-L\`{o}eve, Galerkin, Fokker-Planck. \PACS
89.65.Gh \sep 89.70.+c
\end{keyword}
\end{frontmatter}
Yield curves are remarkable in that the fluctuations of these
structures can be explained largely by a few modes and that the
shape of these modes is largely independent of the market of origin:
a combination of parsimony and explanatory power rarely seen in
financial economics.  While these modes play a fundamental role in
fixed-income analysis and risk management, both the origin of this
modal structure and the relationship between this modal structure
and a formal description of yield curve dynamics remain unclear. The
purpose of this letter is to show that this modal structure is a
natural consequence of the information structure of the yield curve
and that this information structure, in turn, implies an equation of
motion for yield curve dynamics.

Our application of Fisher information to yield curve dynamics is an
extension of prior work~\cite{Frieden98,Frieden04,HawkinsFrieden04a}
using this approach to derive equations of motion in physics and
static probability densities in asset pricing theory\footnote{The
relationship between Fisher information and related approaches such
as maximum entropy~\cite{Jaynes68} and minimum local cross
entropy~\cite{Edelman03} in the context of financial economics is
discussed in~\cite{HawkinsFrieden04a}.}. Though less well known as a
measure of information in physics and mathematical finance than
Shannon entropy, the concept of Fisher information predates
Shannon's and other information statistics, and remains central to
the field of statistical measurement theory~\cite{Kullback59}.
Fundamentally, it provides a representation of the amount of
information in the results of experimental measurements of an
unknown parameter of a stochastic system. Fisher information appears
most famously in the Cramer-Rao inequality that defines the lower
bound on variance/upper bound on efficiency of a parameter estimate
given a parameter dependent stochastic process.  It also provides
the basis for a comprehensive alternative approach to the derivation
of probability laws in physics and other
sciences~\cite{Frieden98,Frieden04,FriedenSoffer00}.

In the present work, the aim of our approach is to derive a
differential equation for yield curve dynamics, {\em ab initio},
with the minimal imposition of prior assumption, save that bond
price observations exist, and that a stochastic process underlies
the dynamics. In a sense our approach is an inversion of the
perspective of a maximum likelihood estimate, where one would solve
for the most likely parameter values given observations within the
context of a pre-assumed model. Here we apply ``minimum presumption"
by deriving the stochastic model that is implied by minimizing
Fisher information given known parameter measurements (bond prices).

A yield curve is a representation of the rate of interest paid by a
fixed-income investment, such as a bond, as a function of the length
of time of that investment.  The interest rate over a given time
interval, between say today and a point of time in the future,
determines the value at the beginning of the time interval of a cash
flow to be paid at the end of the interval; also known as the
present value of the cash flow.  Since the value of any security is
the present value of all future cash flows, yield curve fluctuations
give rise to fluctuations in present value and, thus, play an
important role in the variance of security prices.

The notion that a modal structure underlies yield curve dynamics
comes from common empirical experience with two related yield-curve
measurements - the construction of static yield curves from observed
market prices and the analysis of correlations between corresponding
points in the time evolution of successive static yield curves.
Yield curves are inferred from observed fixed-income security prices
and as the prices of these securities change over time so does the
yield curve.  Yield curves are usually generated after the close of
each trading session and this process can therefore be viewed as a
stroboscopic measurement of the yield curve.  Yield curves can
assume a variety of shapes and many methods have been proposed for
their construction\footnote{See, for example,~\cite{Tuckman02} and
references therein.}.  Of these methods, the Nelson Siegel
approach~\cite{NelsonSiegel87} of representing yield curves as
solutions of differential equations has gained wide acceptance in
the finance industry~\cite{BISYC99} and in academic research on
yield-curve
structure~\cite{Krippner02,DieboldLi02,Krippner03a,Krippner03b,Dieboldetal03}.
In using a second-order differential equation to represent the yield
curve the Nelson Siegel approach is essentially a proposal that
yield curves can be represented effectively by a modal expansion and
the practical success of this approach to yield curve fitting in
markets around the world is a measure of the correctness of this
assertion.

The modal structure of the yield curve is also implied in the
eigenstructure of the two-point correlation function constructed
from yield curves.  Specifically, diagonalization of the covariance
matrix of yield changes along the yield curve produces an
eigenstructure where most of the variance - as measured by summing
the normalized eigenvalues - is explained by the first few
eigenmodes~\cite{Garbade86,LittermanScheinkman91,Garbade96}.  The
consistency of the general shape of the eigenmodes derived from
empirical yield curve data and the explanatory power of the
truncated expansions in those eigenmodes is surprisingly robust over
time and largely independent of the country in which the interest
rates are set~\cite{GarbadeUrich88,Phoa98,Phoa00}.  While this
analysis motivated the use of yield-curve modes by fixed-income
strategists and risk managers some time ago, an explicit link
between yield-curve modes and dynamics appeared in comparatively
recently research demonstrating the eigenstructure to be consistent
with both the existence of a line tension along the yield curve and
a diffusion-like equation of motion for yield
curves~\cite{Bouchaudetal99}.  This notion of a line tension along
the yield curve has found further expression in descriptions of the
yield curve as a vibrating string~\cite{SantaClaraSornette01}. While
the yield curve phenomenology just described {\em can} be described
well by modal expansions there as been little to motivate {\em why}
this should be the case and it is to this question that we now turn.

We begin with a more formal description of the notion of present
value mentioned above.  The yield curve is closely related to the
function $D(t,T)$ known as the discount function that gives the
value at time $t$ (i.e. the present value) of a unit of currency
(e.g. dollar) to be paid at time $T$ in the future
\begin{equation}
D(t,T) = e^{-r_s(t,T)(T-t) } = e^{ -\int_t^{t+T} r_f(t,s) ds } \; ,
\label{eq:df-yc}
\end{equation}
where $r_s(t,T)$ is the ``spot rate'' yield curve at time {\em t},
specifying the continuously compounded interest rate for borrowing
over the period $\left [ t, T \right ]$, and $r_f(t,T)$ is known as
the ``forward rate'' yield curve at time {\em t}, specifying the
interest rate for instantaneous borrowing over $\left [ T, T+\delta
T \right ]$.

Our explanation for the existence of dynamic yield-curve modes
builds on our recent application of Fisher information
methods~\cite{Frieden04,HawkinsFrieden04a} to the construction of
well-mannered, static yield curves from a finite set of discount
functions or observed bond prices.  Our approach, based on deriving
yield curves that extremize Fisher information, is facilitated by
associating such yield curves with complementary probability density
functions where the time to maturity $T$ is taken to be an abstract
random variable~\cite{BrodyHughston01,BrodyHughston02}.  We assume
the associated probability density $p(t,T)$ satisfies $p(t,T) > 0$
and is related to the discount factor $D(t,T)$ via
\begin{equation}
D(t,T) = \int_T^{\infty} p(t,s) ds \; . \label{eq:df}
\end{equation}
Discount factors are, however, not always observable.  Coupon bonds,
on the other hand, are commonly available with prices $B(t,N)$
related to the discount factor by
\begin{equation}
B(t,N) = \sum_{i=1}^N C \left ( T_i \right ) D \left ( t, T_i \right ) \; ,
\label{eq:cb}
\end{equation}
where $N$ indicates the number of remaining coupon payments and  $C
\left ( T_i \right )$ is the cash flow at time $T_i$ in the future.
For a typical coupon bond $C \left ( T_i \right )$ is equal to the
$i^{th}$ coupon payment for $i < N$ and equal to the final coupon
payment plus the principal payment for $i = N$.

In these expressions one can see that discount factors and bond
prices share a common structure as averages of known functions.
Discount factors are the average of $\Theta \left ( s - T \right )$
and coupon bond prices are the average of $\sum_{i=1}^N C \left (
T_i \right ) \Theta \left ( s - T_i \right )$ where $\Theta(x)$ is
the Heaviside step function.  Generally, where observed data $d_1,
\ldots , d_M = \{d_m\}$ such as discount factors and the prices of
bonds can be expressed as averages of known functions $\{f_m\}$ at a
static point in time, we may write
\begin{equation}
\int f_m(T) p(T)dT = d_m \; , \; \; \; m = 1, \ldots , M \label{eq:constraints}
\end{equation}
and the probability density function $p(T)$ implicit in the observed
data can be obtained by forming a Lagrangian using Fisher
information~\cite{Fisher25} in its shift-invariant form
\begin{equation}
I = \int dT \frac{\left ( dp(T)/dT \right )^2}{p(T)} \; .
\end{equation}
Employing the usual variational approach we obtain~\cite{Frieden04}
\begin{eqnarray}
p(T) & = & q^2(T) \; , \\
\frac{d^2 q(T)}{dT^2} & = & \frac{q(T)}{4} \left [ \lambda_0 + \sum_{m=1}^M \lambda_m f_m(T) \right ] \; , \label{eq:swe}
\end{eqnarray}
where the $\lambda$'s are Lagrange multipliers that enter by
incorporating a normalization constraint on $p(T)$ ($\lambda_0$) and
observed data ($\lambda_m$) into the Fisher information Lagrangian.

This is equivalent to an approximate use of the extreme physical
information (EPI)
approach~\cite{Frieden98,Frieden04,FriedenSoffer00} wherein the
constraint equations (Eq.~\ref{eq:constraints}) are used in place of
the fundamental source information {\em J} required by EPI.  This
replacement amounts to a technical (as compared with fundamental)
approach to valuation.

\begin{figure}
\includegraphics[scale=0.55]{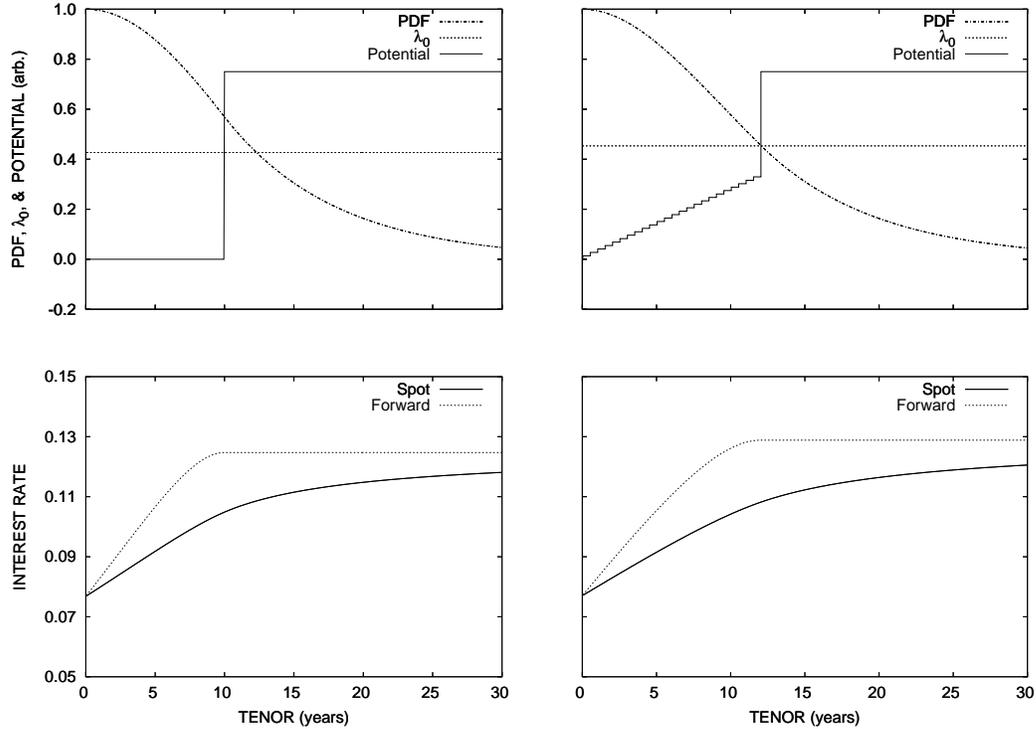}
\caption{The equilibrium densities, related functions, and implied
yield curves for a discount factor (left) and a coupon bond
(right).} \label{fig:ground}
\end{figure}

In a recent communication~\cite{HawkinsFrieden04a} we exploited the
formal equivalence of Eq.~\ref{eq:swe} and the time-independent
Schroedinger wave equation (SWE) to calculate the equilibrium
densities $p(T)$ implicit in security prices as shown in
Fig.~\ref{fig:ground}.
In the graphs on the left-hand side of Fig.~\ref{fig:ground} we see
the results for a single discount factor with a price of 35\% of par
and a tenor of 10 years.  The upper-left graph illustrates the
elements of Eq.~\ref{eq:df} with the potential term of the SWE
($\sum_{i=1}^M \lambda_m f_m(T)/4$) being a single step function.
The amplitude of the potential well described by the step function,
$\lambda_1$, the level of $\lambda_0$ and the probability density
function (PDF) $p(T)$, all follow from a self-consistent field
calculation with the $\{\lambda_0, p(T)\}$ pair corresponding to the
ground state of Eq.~\ref{eq:swe} subject to the constraint given by
Eq.~\ref{eq:df}.  The lower-left graph shows the spot and forward
yield curves that follow from the PDF in the upper graph as defined
by Eq.~\ref{eq:df-yc}.

The results of this analysis for a 6.75\% coupon bond making
semi-annual payments with a maturity date of November 15, 2006, a
price of 77.055\% of par, and a pricing date of October 31,
1994~\cite{FrishlingYamamura96} are shown on the right-hand side of
Fig.~\ref{fig:ground}.  The stepped structure of the potential
function is a result of the cumulative sum of the coupon payments
with the final large step being due to the principal payment. Unlike
the discount factor, there is no analytic solution to
Eq.~\ref{eq:swe} for the coupon bond.  This type of potential is,
however, ideally suited to the transfer matrix method of
solution~\cite{Yehetal77} and that is the approach we used to
calculate the PDF solution shown in the upper-right graph.  The
calculation of a general yield curve from a collection of coupon
bonds is a straightforward extension of this approach.

The general solution to the SWE with potentials of this form is
commonly expressed as a series expansion of modes and these modes
have been used to go beyond the equilibrium solutions illustrated in
Fig.~\ref{fig:ground} to describe {\em non-equilibrium} phenomena in
physics~\cite{Friedenetal02a,Friedenetal02b,Flegoetal03}.  It is
with these modes that the modal structure of the yield curve follows
directly from the Fisher information structure of the yield curve
(cf. Eq.~\ref{eq:prop} below).  This result also provides an
information-theoretic derivation of the Nelson Siegel approach.  Of
interest as well is the behavior of the solutions illustrated in
Fig.~\ref{fig:ground} in the range of tenor where there are no
observed security prices.  The solution to Eq.~\ref{eq:swe} is known
to be an exponential decay which leads to a constant interest rate:
a result consistent with most priors concerning long-term interest
rates.

The temporal evolution of yield curves now follows directly from the
known relationship between solutions of Eq.~\ref{eq:swe} and those
of the Fokker -Planck equation\footnote{Formally, the Fokker-Planck
equation can be obtained from our Fisher Information based
variational approach by incorporating a Lagrangian term enforcing
the constraint that total probability density is conserved under
time
evolution~\cite{ReginattoLengyel04}.}\cite{VanKampen77,Risken96}.
Specifically, the solutions of Eq.~\ref{eq:swe} $\{\lambda_0^{(m)},
q_m(T)\}$ can be used to construct a general
solution~\cite{VanKampen77}
\begin{equation}
p(T,t) = \sum_{m=0}^{\infty} c_m q_0(T) q_m(T) e^{-\vartheta \left ( \lambda_0^{(m)} - \lambda_0^{(0)} \right ) t/4} \label{eq:prop}
\end{equation}
to the Fokker-Planck equation
\begin{equation}
\frac{\partial p(T,t)}{\partial t} = \frac{\partial}{\partial T} \left [ \frac{\partial U(T)}{\partial T} + \vartheta \frac{\partial}{\partial T} \right ] p(T,t) \; , \label{eq:fpe}
\end{equation}
where the potential function $U(T)$ is related to the ground state $q_0(T)$ via
\begin{equation}
U(T) = -2 \vartheta \log q_0(T) \; . \label{eq:propend}
\end{equation}
Taken together, Eqs.~\ref{eq:swe} through~\ref{eq:propend} explain
the existence of a modal structure of yield curves and provide a
theoretical basis for the common ansatz that a diffusion process
underlies interest-rate
dynamics~\cite{Tuckman02,Garbade96,Rebonato98,BrigoMercurio01}.

There are a variety of ways to solve Eq.~\ref{eq:prop}, but the
observation that the the eigenstructure of the two-point correlation
function is dominated by a few modes suggests that this infinite
series can be reduced to a few terms using the Karhunen-L\`{o}eve
expansion\footnote{This approach appears under a variety of names
including factor analysis, principal-component analysis, and proper
orthogonal decomposition.} together with the Galerkin
approximation~\cite{Sirovich87a,Sirovich87b,Sirovich87c,BreuerSirovich91}.
Specifically, writing Eq.~\ref{eq:prop} in the slightly more general
form $p(T,t) = \sum_{m=0}^{\infty} a_m(t) \phi_m(T)$ where
$\phi_m(T) \equiv q_0(T)q_m(T)$, substituting this into the
Fokker-Planck equation written suggestively as $\dot{p} =
\mathcal{L}_{FP}(p)$, and projecting along the eigenfunctions
$\phi_m(T)$ one obtains
\begin{eqnarray}
\dot{a}_i(t) & = & \int_0^{\infty} \mathcal{L}_{FP} \left ( \sum_{m=0}^{\infty} a_m(t) \phi_m(T) \right ) \phi_i(T) dT \; , \\
a_i(0) & = & \int_0^{\infty} p(T,0) \phi_i(T) dT \; .
\end{eqnarray}
Truncating the series expansion for $p(T,t)$ at $i=N$ gives a Galerkin approximation of order $N$~\cite{Sirovich87c} and this truncation is justified in our case because of the dominance to the two-point correlation function by a few modes.
\begin{figure}
\includegraphics[scale=0.55]{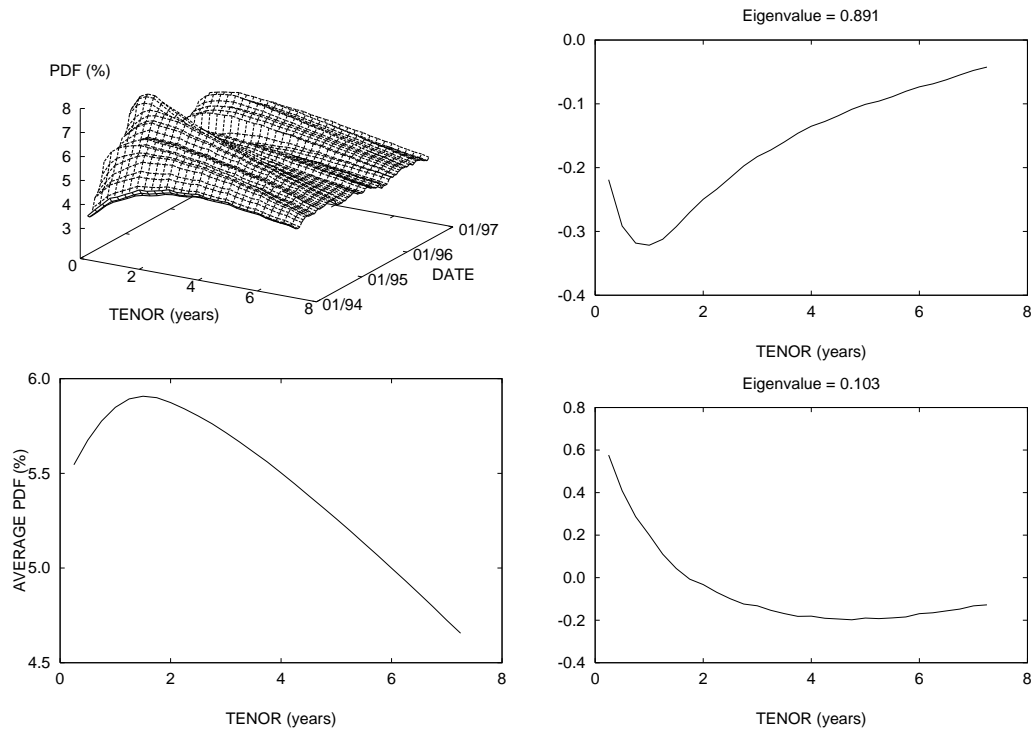}
\caption{The PDF as a function of time for Eurodollar futures (upper left) together with the average PDF during this period (lower) left and the empirically determined eigenfunctions with corresponding normalized eigenvalues (upper and lower right).}
\label{fig:ed}
\end{figure}

An example of applying this approach to the dynamics of the
Eurodollar yield curve is illustrated in Fig.~\ref{fig:ed}.  The
probability density function implicit in the Eurodollar futures
market from the beginning of 1994 to the end of 1996 is shown in the
upper-left panel of the figure and the average of these density
functions is shown in the lower-left panel\footnote{The probability
density function was obtained from constant-maturity Eurodollar
futures prices as discussed in~\cite{Bouchaudetal99}}.  Using the
method of snapshots~\cite{Sirovich87a} we obtained the
eigenstructure shown partially in the two right-hand panels of
Fig.~\ref{fig:ed}.  As the normalized eigenvalues indicate, more
than 99\% of the variance of this system is contained in the first
two modes.  Thus a Galerkin approximation of order 2 would be
expected to provide an adequate representation of yield curve
dynamics.

In summary, we have derived an equation of motion for yield curves
that is consistent with observed statics and dynamics starting from
the Fisher information of the probability density function that
underlies the discount function.  Our derivation leads to a
Schroedinger wave equation for the probability amplitude of the
density function underlying the discount factor and thus explains
why solutions of equations of mathematical physics involving
second-order tenor derivatives work so well as a representation of
yield curves.  This result also provides an explanation for the
existence of a line-tension term in the equations of motion found in
string models of yield curves.  Using the well-known relationship
between solutions of the Schroedinger wave equation and the
Fokker-Planck equation we obtained an equation of motion for the
yield curve consistent with the common ansatz that diffusion
processes underly yield-curve dynamics.  Since the eigenstructure of
the yield-curve two-point correlation function is dominated by a few
modes we found that a practical numerical solution to this equation
of motion can be had by using the Karhunen-L\`{o}eve expansion
together with Galerkin's method.

\end{document}